\documentclass[aps,prapplied,twocolumn,superscriptaddress]{revtex4-2}

\usepackage{amsmath, amssymb, amsthm, amsfonts, latexsym, graphicx, mathtools}
\usepackage{float}
\usepackage{verbatim}
\usepackage{hyperref}
\usepackage{dsfont}
\usepackage{wrapfig}
\usepackage{mathrsfs}
\usepackage[caption=false]{subfig}
\usepackage{cleveref}
\usepackage[normalem]{ulem}
\usepackage{braket}
\usepackage{ragged2e}
\usepackage[dvipsnames]{xcolor}
\usepackage{booktabs}
\hypersetup{
  colorlinks   = true,
  urlcolor     = blue,
  linkcolor    = blue,
  citecolor    = blue
}

\DeclareMathOperator*{\argmin}{argmin}

\begin{document}
\title{\textbf{Machine Learning Interpretability in Photonics: Optimizing Inverse Design for Enhanced Device Performance}}
\title{\textbf{Enhanced Photonic Chip Design via Interpretable Machine Learning Techniques}}
\author{Lirandë Pira}
\email{lpira@nus.edu.sg}
\affiliation{Centre for Quantum Technologies, National University of Singapore, Singapore}
\author{Airin Antony}
\email{airin.antony@uq.edu.au}
\affiliation{School of Mathematics and Physics, University of Queensland, Australia}
\author{Nayanthara Prathap}
\affiliation{Centre for Quantum Technologies, National University of Singapore, Singapore}
\author{Daniel Peace}
\affiliation{School of Mathematics and Physics, University of Queensland, Australia}
\author{Jacquiline Romero}
\affiliation{Australian Research Council Centre of Excellence for Engineered Quantum Systems, Australia}
\affiliation{School of Mathematics and Physics, University of Queensland, Australia}

\begin{abstract}
Photonic chip design has seen significant advancements with the adoption of inverse design methodologies, offering flexibility and efficiency in optimizing device performance. However, the black-box nature of the optimization approaches, such as those used in inverse design in order to minimize a loss function or maximize coupling efficiency, poses challenges in understanding the outputs. This challenge is prevalent in machine learning-based optimization methods, which can suffer from the same lack of transparency. To this end, interpretability techniques address the opacity of optimization models. In this work, we apply interpretability techniques from machine learning, with the aim of gaining understanding of inverse design optimization used in designing photonic components, specifically two-mode multiplexers. We base our methodology on the widespread interpretability technique known as local interpretable model-agnostic explanations, or LIME. As a result, LIME-informed insights point us to more effective initial conditions, directly improving device performance. This demonstrates that interpretability methods can do more than explain models --- they can actively guide and enhance the inverse-designed photonic components. Our results demonstrate the ability of interpretable techniques to reveal underlying patterns in the inverse design process, leading to the development of better-performing components. 
\end{abstract}
\date{\today}
\maketitle

\newpage

\section{Introduction}\label{sec:introduction}
Photonic chip design has seen remarkable progress in the past years due to advancements in both materials science and fabrication techniques \cite{obrien2003controllednot, pooley2012controllednot, Paesani_2020, katsumi2019quantumdot, romero_photonicquantumcomputing_2024}. The development of photonic materials, such as silicon photonics and two-dimensional materials like graphene, has contributed to better performance and scalability of photonic chips. At the same time, this progress is aided by advances in fabrication methods, such as automated design tools. A recent development that is gaining popularity is the concept of inverse design, a paradigm shift from traditional, intuition-driven design methodologies to algorithmically optimized designs. While inverse design has demonstrated the boundaries of photonic chip performance, its widespread adoption is, in part, hindered by the opacity of the optimization process. Existing inverse design methodologies rely on optimization methods, which operate as black-boxes. This optimization process yields component designs that are not always intuitive, hindering the researcher's ability to understand and improve the device.
\begin{figure*}[t]
    \centering
    \includegraphics[width=0.99\linewidth]{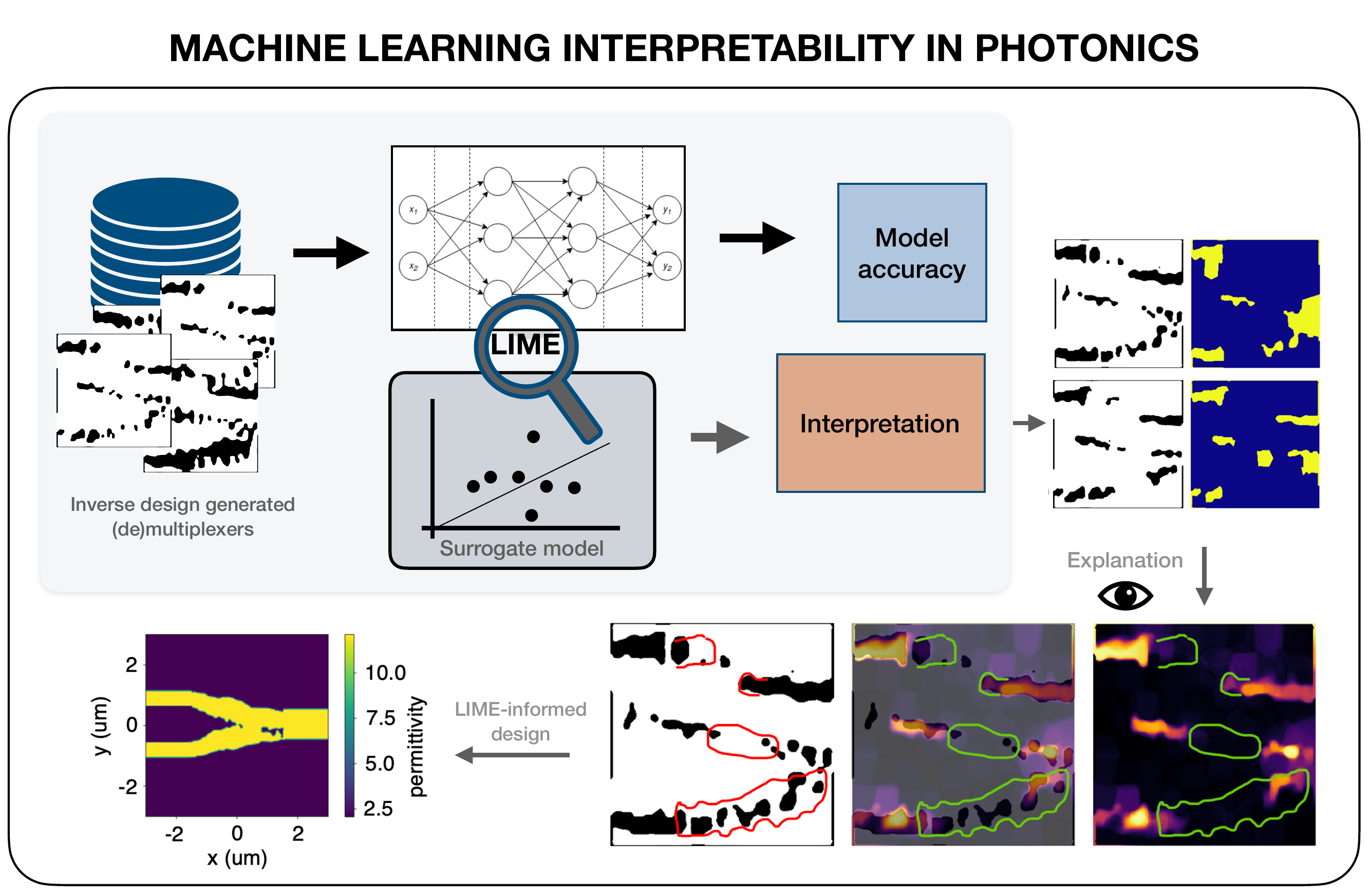}
    \caption{\justifying Illustration of the LIME process integrated in the interpretation of inverse design outputs. Dataset is a collection of inverse design generated samples. Neural network represents a non-linear often complex model architecture which outputs a certain metric of choice, such as prediction accuracy. Function surrogate model is a simpler model, such as a type of regression applied locally on the level of a data sample. The output of the surrogate model provides insights into the predictions of the neural network.}
    \label{fig:main}
\end{figure*}

Similarly, machine learning (ML) models optimize the process of learning from data \cite{mitchell_machine_1997, bishop_pattern_2006}. ML models deployed for large scale systems are powered by neural networks, which are highly non-linear functions, and thus behave as opaque systems. Their decision-making processes are difficult to interpret or explain, even for experts. This lack of transparency into the decision-making process of these models, gives rise to the concept of interpretability or explainability, whereby interpretable methods or interpreters are developed. Interpretability in machine learning is about how well one can understand a model's predictions and decisions \cite{molnar_interpretable_2022, burkart_survey_2021}. It includes transparency, where simpler models are often easier to grasp than complex ones, and explainability through methods like LIME (local interpretable model-agnostic explanations) \cite{ribeiro_why_2016} and SHAP (Shapley additive explanations) \cite{lundberg_unified_2017} that highlight which features in a data space matter most. In general, this understanding of model outputs is important for several reasons. First, it helps in spotting biases and improving models, and secondly, it helps model debugging and building trust in model outputs.

In response to the challenge of opacity in photonic optimization methods, in this work, we integrate interpretable machine learning techniques into the workflow of inverse design for photonic components. These interpretable methods aim to provide more insights into model decisions, revealing the intricate relationships between device geometries and their performance. By doing so, we open the optimization black-box behind these models, which in turn, allows for building better photonic components and devices. Here, the interpreter of choice is LIME. Briefly, LIME belongs to a class of interpreters known as surrogate models, where a simpler model, such as linear regression, is used to explain the outputs of a more complex one, such as a convolutional neural network. We apply the interpreter to a commonly-used photonic component, the mode (de)multiplexer, to study its bandwidth characteristics. We then use the resulting insights to choose a suitable initial condition for the inverse design optimization which improves the final device performance. Our findings suggest potential for further exploration. A high level depiction of this process is illustrated in \Cref{fig:main}.

This work demonstrates how interpretable machine learning, specifically LIME, can provide actionable insights into photonic device performance --- in this case, bandwidth of inverse-designed multiplexers. Analysis of LIME heatmaps overlaid on device geometries revealed that structural features such as gaps and small etched regions correlate with lower bandwidth. These observations were substantiated by finite-dimensional time-domain (FDTD) simulations, which indicated that such features contribute to increased optical leakage and scattering. Informed by these insights, we introduced targeted manual design modifications and refined initial conditions for inverse design, resulting in devices with markedly improved bandwidth and transmission. Our approach not only improved the efficiency of the design process, but also demonstrated that explainable AI tools can meaningfully influence photonic design strategies.

Similar ideas on interpreting inverse design outputs are explored, focusing on how machine learning models can be applied to enhance the understanding of complex design solutions in different fields.
For example, Ref.~\cite{zhu2022harnessing} uses an interpretable method called decision tree-random forest to understand inverse design in origami-type structures in engineering, by building four datasets based on four design features. Ref.~\cite{jia2023interpretable} proposes a new approach to interpretation-friendly inverse design in photonics, specifically for nonlinear systems such as four-wave mixing, which have been underexplored compared to linear systems. The study demonstrates the proposed method by fabricating a compact device on a silicon-on-insulator platform that generates photon pairs at a rate of $1.1$ MHz. Ref. ~\cite{chen2024generative} introduces a generative inverse design technique that uses an interpretable random forest-based forward model to map design to the response. This eliminates the need for a complex inverse model. Here, interpretability is an inherent part of the learning model, rather than an added feature or post-processing step. The highlighted advantages of this method are twofold: interpretability, and the ability to remain efficient under small data constraints. Ref.~\cite{zeng2024dataefficientinterpretableinversematerials} introduces a semi-supervised learning method using a disentangled variational autoencoder for inverse materials design, which separates target properties from other material features, making the model more transparent. Additional interpretability is provided through post-hoc analysis of the model's classification layer.

This manuscript is structured as follows. \Cref{sec:integrated-photonics} introduces integrated photonics and mode (de)multiplexers. \Cref{sec:inversedesign} focuses on photonic inverse design with a focus on how these techniques can be applied to optimize photonic devices at the chip level. \Cref{sec:mlinterpretability} introduces supervised machine learning, neural network-based learning models, and the ML interpretability technique called LIME. \Cref{sec:numerical_methods} outlines the numerical methods, including the generation and preprocessing of the (de)multiplexer dataset, model setup, training procedure, and the LIME methodology. It details the design parameters, performance metrics, and data preparation steps for machine learning. Findings from numerical results are discussed in \Cref{sec:numerical-results}. These findings are applied to choose a better initial condition for the inverse design method in generating (de)multiplexers, detailed in \Cref{sec:initial-condition}. Further insights into (de)multiplexer generation process are outlined in \Cref{sec:mux-generation}. This study concludes in \Cref{sec:conclusion} with a broader discussion of the implications of our work and proposed potential directions for future research.

\section{A Brief Introduction to Integrated Photonics}\label{sec:integrated-photonics}
\subsection{Silicon Integrated Photonics}
A photonic integrated circuit uses photons as the main carrier of information. Components for manipulating photonic properties can be integrated into stable and compact on-chip photonic circuits that find applications in many different areas, i.e., lidar systems, telecommunications, environmental and biomedical sensing, quantum computing, and optical signal processing. Different platforms can be used, i.e., silicon, silicon nitride, indium phosphide and lithium niobate, with each platform showcasing distinct optical properties and potential for scalability \cite{butt2025lighting}.

Among different material platforms, silicon integrated photonics is especially attractive due to its compatibility with the existing semiconductor fabrication techniques which offers scalability \cite{Jalali2006Silicon, bogaerts2018silicon}.
Silicon is transparent between $1.1$ $\mu$m to 4 $\mu$m, which include the range most relevant to telecommunications. For a detailed review of silicon photonics, refer to Siew \textit{et}. \textit{al}. \cite{siew_review_2021}. In this work, we employ silicon photonics to illustrate our results; however, the methodology presented can be readily applied to photonic components in other platforms as well. Photonic components are the building blocks used in any photonic system. The role of these components is to manipulate and utilize photons for various applications. These components are typically integrated into photonic circuits, often on a chip-scale platform, to perform specific functions such as signal transmission, switching, modulation, and processing. In this paper, we focus on a key component in photonic designs, the mode multiplexer.
 
\subsection{Mode (De)Multiplexers}
In optical systems, light can propagate in different modes based on parameters such as the waveguide geometry, the wavelength of light, and its polarization. Common types of polarization modes include transverse electric (TE) modes and transverse magnetic (TM) modes. TE/TM modes have electric/magnetic fields that are entirely transverse to the direction of propagation. Mode multiplexers/demultiplexers generally refer to components that combine/separate multiple light modes or different spatial or polarization modes into/from a single waveguide. The same photonic component can act as a multiplexer or demultiplexer depending on the propagation direction. In this work, mode (de)multiplexers refer specifically to TE/TM (de)multiplexers. These components are integral to many emerging applications, such as higher-order spatial switches, reconfigurable mode-selective switches, programmable optical processors, and multimode quantum photonics \cite{mojaver2024recent}. They help improve intra-chip and chip-to-chip communication, enhancing data transmission throughput and network flexibility \cite{sun2025edge, mojaver2024recent}.

\section{Inverse Design in Integrated Photonics}\label{sec:inversedesign}
\subsection{Motivation for Inverse Design}
Traditionally, photonic components in silicon and other platforms were optimized by tuning a handful of basic parameters, i.e., dimensions of waveguides, gap between adjacent structures, etc. These bottom-up approaches, also known as the forward design paradigm, are mostly intuition driven. Numerous photonic components, such as waveguides, modulators, and filters, have been designed using these techniques.

However, standard fabrication processes offer a minimum feature size of tens of nanometers, meaning that thousands of degrees of freedom are left unused by conventional methods. The resulting components are bulky, less efficient, and often tedious to optimize further. Additionally, photons enable high-dimensional information processing --- the signals can be encoded in hundreds of wavelengths, time bins, multiple waveguide paths, and a handful of TE/TM modes, or even in some combination of these. The effects of under-utilized parameters scale up quickly when these diverse photon properties are involved.

For instance, consider the case of directional couplers, a basic component that can transfer light across neighboring waveguides. These structures can act as mode (de)multiplexers by converting one transverse mode to another across neighboring waveguides. Optimizing them while maintaining compactness is hard since there are only a handful of parameters to tune --- the lengths and widths of the waveguides and the separation between them. Even a basic TE0-TE1 or TE0-TE2 directional coupler can have a coupling length of roughly $30$ $\mu$m \cite{yin2023integrated}; the length of a multi-mode (de)multiplexer will scale linearly with number of modes. This does not account for additional components in the device, such as waveguide bends and adiabatic waveguide regions, which can range from tens to hundreds of micrometers. Devices implementing more complex operations tend to become increasingly bulky and lossy.

The growing demands of photonic circuits are enabling a paradigm shift away from such conventional methods.
Non-intuitive, machine learning-based techniques like inverse design promise more compactness, bandwidth and performance. In the context of mode-(de)multiplexing discussed above, consider a recent result in this space that is pushing the boundaries of high-capacity photonic interconnects: an inverse-designed five-mode demultiplexer that spans only $10 \times 6 \mu$m$^{2}$ \cite{sun2025edge}. Below, we discuss this growing field of photonic inverse design in greater detail.

\subsection{Nanophotonic Topology Optimization}
Access to exponentially more computing power and powerful optimization tools in the recent decades have allowed researchers to manipulate significantly more design parameters. Such optimization techniques have enabled devices with improved performance in terms of compactness, bandwidth, fabrication robustness, and efficiency. For a comprehensive list of nanophotonic devices designed using optimization algorithms, we refer the reader to Park \textit{et.} \textit{al.} \cite{park_free_2022}; the algorithms explored include principal component analysis \cite{melati_mapping_2019}, genetic algorithms \cite{yu_genetic_2017}, adjoint-based methods \cite{jensen_systematic_2004, wang_robust_2011}, and deep neural networks \cite{tahersima_deep_2019}, among others.

\begin{figure*}[t]
    \centering
    \includegraphics[width=0.85\linewidth]{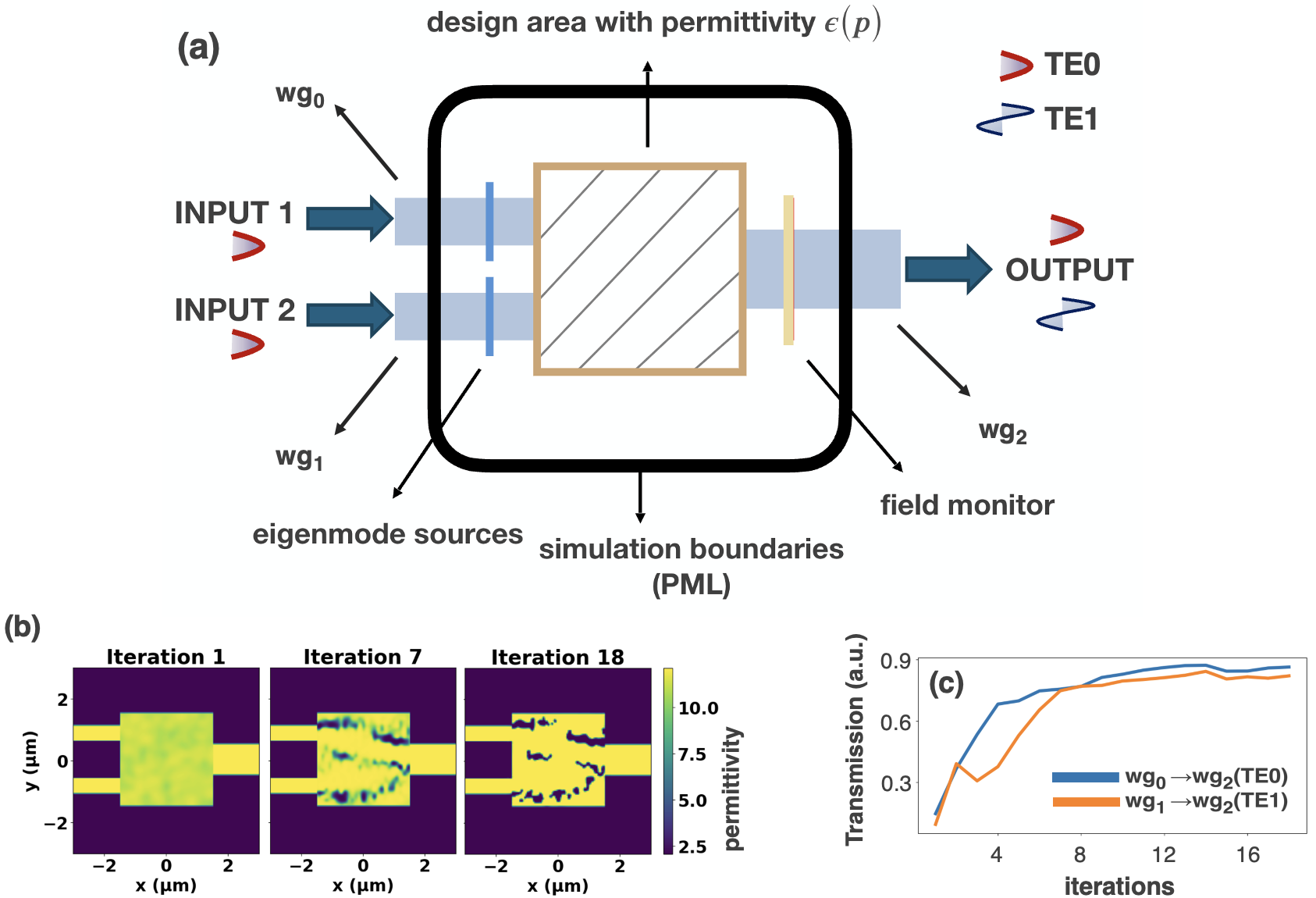}
    \caption{\justifying a) Schematic representation of a photonic inverse design simulation. Design region is characterized by a permittivity distribution $\epsilon(p)$. The system features two input channels modeled as eigenmode sources (wg$_{0}$(TE$0$) and wg$_{1}$(TE$0$) modes), and one output channel (supporting wg$_{2}$(TE$0$) and wg$_{2}$(TE$1$) modes). The simulation boundaries use perfectly matched layer (PML) conditions to minimize reflection. An output field monitor evaluates performance at each iteration. (b) A mode multiplexer at various stages of the inverse design optimization. The objective function prioritized maximizing transmissions from wg$_{0}$(TE$0$) to wg$_{2}$(TE$0$) and from wg$_{1}$(TE$0$) to wg$_{2}$(TE$1$) at the wavelength of $1.55~\mu$m. The transmissions improving over the course of optimization is shown in plot (c).}
    \label{fig:fig-inverse}
\end{figure*}

Of particular interest is adjoint-based topology optimization, a design-by-specification method capable of designing various functionalities such as wavelength and mode multiplexers, splitters, and couplers \cite{kontoleontos2013adjoint, papoutsis2016continuous}.
Here, the user defines the inputs and the desired operation. Starting from a random parameterized permittivity distribution, the inverse design algorithm finds the optimal design within the given material and fabrication constraints.
The goal is to minimize the user-specified objective function $f_{obj}$:
\begin{equation}
    min~f_{obj} (p) = (t - |c^{\dagger}E(\epsilon(p))|)
^2
\end{equation}
where $t$ is the target transmission, $\epsilon$ is the permittivity distribution parameterized by $p$ which accounts for fabrication constraints, and $c^{\dagger}E$ the modal overlap between target mode $c$ and electric field $E$ at the output field monitor. A schematic of this process is illustrated in \Cref{fig:fig-inverse}. Multiple objective functions may be specified, accounting for cross-talk, bandwidth, temperature-sensitivity, or fabrication-constraints \cite{su_nanophotonic_2020}.
The gradient at each step is calculated from the forward and adjoint electromagnetic simulations using reverse-mode auto-differentiation \cite{su_nanophotonic_2020, lee_systematic_1997}, and the design parameters are updated so as to minimize the objective function.
Here, we use the adjoint-based inverse design package, Stanford photonic inverse design software (SPINS) \cite{su_nanophotonic_2020}. SPINS uses finite-difference frequency-domain (FDFD) electromagnetic simulations to obtain the fields for each iteration. At the start, the user specifies the material properties, design area, minimum feature size, simulation boundaries, and the objective function. The software uses a bicubic interpolation to initialize a randomized continuous permittivity distribution function over the design area which then gets optimized. A sigmoid factor controls the discreteness or the steepness of the curves, which gradually increases. Afterwards, the final design is discretized using the levelset method and further optimized by levelset parameterization.

\subsection{Opacity of the Inverse Design Process}
\begin{figure}[t]
    \centering
    \includegraphics[width=0.99\linewidth]{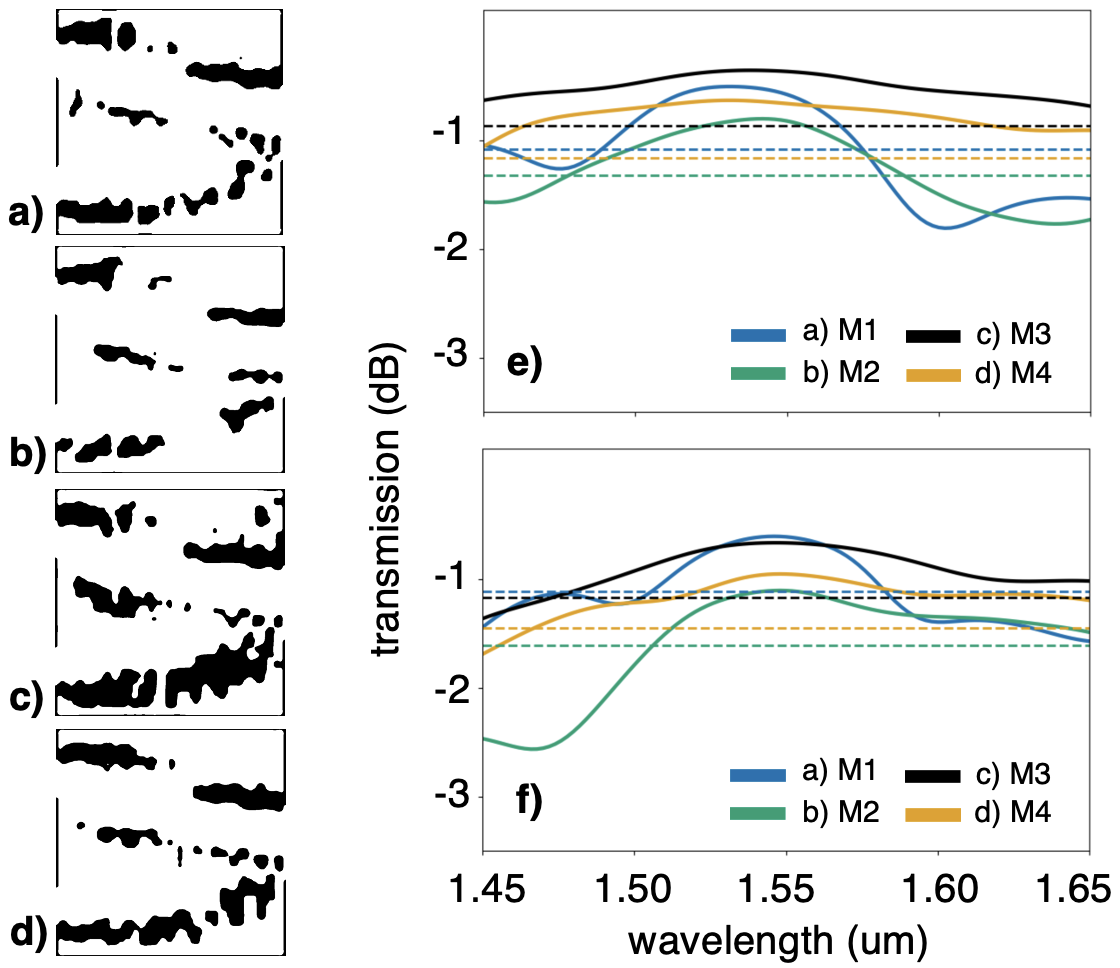}
    \caption{\justifying Inverse designed multiplexers M1-M4 in grayscale (a-d), and their transmission spectra for Mode~1 (e) and Mode~2 (f). The dotted lines show the corresponding $0.5$ dB bandwidth cut-offs. M3 and M4 have much higher bandwidths even though similar design parameters were used.}
    \label{fig:fig-opacity}
\end{figure}
Despite the disadvantages of conventional forward designs, their intuitive nature is valuable for physicists --- for the sake of understanding how a given component works (or does not work as expected), and to improve designs in a predictable fashion. With inverse designed components however, there exists an inherent transparency issue. Consider the four inverse designed mode-multiplexers (M1 to M4) in \Cref{fig:fig-opacity}. The devices were designed using SPINS by optimizing for transmissions from wg$_{0}$(TE$0$) $\rightarrow$ wg$_{2}$(TE$0$) (Mode~1) and from wg$_{1}$(TE$0$) $\rightarrow$ wg$_{2}$(TE$1$) (Mode~2), at wavelength $1.55$ $\mu$m. A total of 18 iterations were used and the design parameters were kept the same for all devices. We plot the transmission spectra using FDTD solver from Tidy3D software \cite{flexcompute2022tidy3d}.

At first glance, devices M1-M4 seem to share some similarities. Their transmissions seem to peak around $1.55$ ${\mu}$m and vary from -0.4 to -1.1 dB. The bandwidths are much higher than conventional designs as expected. However, on closer examination, the bandwidths of devices M$1$ and M$2$ are noticeably lower than the other two ($84-110$ nm vs $>200$ nm for Mode~1, and $76-144$ nm vs $176-182$nm for Mode~2). The design parameters are equal; there is no reliable way to discern which features contributed to the relatively higher bandwidths of M3 and M4. The only viable ways to improve the design in terms of bandwidth would be to blindly iterate the process by varying the design parameters, or to add additional constraints on the objective function by considering multiple wavelengths (which also increases computational resources). This black-box nature of inverse design hinders our attempts to understand and improve the model or to provide a satisfactory reasoning for the design parameters and initial conditions used. In this context, techniques that can provide insight into the inverse design component and highlight significant features can be valuable. Our work focuses on the case of mode multiplexers and their bandwidth characteristics.

\section{Machine Learning and Interpretability Methods}\label{sec:mlinterpretability}
\subsection{Supervised Learning Models}\label{subsec:supervisedlearning}
Machine learning involves training models to classify data and make predictions based on patterns learned from examples. This process can take various forms, depending on the type of data, the learning objectives, and the techniques used. One of the most widespread ideas is the notion of supervised learning where a model is trained on a labeled dataset $D = \{(x_i, y_i)\}_{i=1}^n$ consisting of $n$ samples. Each input sample $x_i \in \mathbb{R}^d$ is associated with a corresponding label $y_i$, which can be categorical for classification tasks or continuous for regression tasks. The objective of supervised learning is to learn a mapping function $f: \mathbb{R}^d \to \mathbb{R}$ that approximates the true relationship between input and output samples. To achieve this, we define a loss function $L(y, f(x; \theta))$ that quantifies the error between the predicted output $f(x; \theta)$ and the true label $y$, where $\theta$ represents the model parameters.

Learning process involves minimizing the empirical risk, or training error, expressed as
\begin{equation}
    \theta^* = \arg \min_{\theta} \frac{1}{n} \sum_{i=1}^n L(y_i, f(x_i; \theta)).
\end{equation}

This optimization is typically carried out using algorithms such as gradient descent, which iteratively updates the parameters according to the rule $\theta \leftarrow \theta - \eta \nabla_{\theta} J(\theta)$, where $\eta$ is the learning rate and $J(\theta)$ is the average loss over the training set. After training, the model's performance is evaluated on a separate test set $D_{test} = \{(x_j, y_j)\}_{j=1}^m$ to assess its ability to generalize, quantified by the test error expressed as
\begin{equation}
    \mathcal{E}_{test} = \frac{1}{m} \sum_{j=1}^m L(y_j, f(x_j; \theta^*)).
\end{equation}

Neural network-based learning models are a class of ML algorithms designed to approximate complex, non-linear mappings between input features and output predictions. At their core, neural networks consist of layers of interconnected artificial neurons, with each neuron performing a weighted sum of its inputs followed by a non-linear activation function, such as ReLU, sigmoid, or tanh \cite{goodfellow_deep_2016}. Training a neural network involves optimizing a loss function, typically using gradient-based methods such as stochastic gradient descent or its variants (i.e., Adam, RMSprop), a process highlighted in \Cref{subsec:supervisedlearning}.

The model employed in the numerical experiments is a convolutional neural network (CNN), which is a type of deep learning model designed for processing grid-like data, such as the \emph{bmp} images \cite{LeCun1998, oshea2015introductionconvolutionalneuralnetworks}. The reason why they perform well on tasks in computer vision, is because they are designed to adaptively learn spatial hierarchies of features from data. More specifically, CNNs consist of several key layers that work together to extract and process features from input data. The convolutional layers serve as the core building blocks, where each layer applies a set of filters (or kernels) that slide over the input, performing convolution operations to generate feature maps highlighting different aspects of the data. Following these, pooling layers reduce the spatial dimensions of the feature maps, effectively down-sampling the data to retain essential information while improving computational efficiency. Finally, fully connected layers appear at the end of the network, combining the extracted features to produce the final output, such as predictions or classifications.

In addition, similar to other types of neural network architectures layers which introduce non-linearity through activation functions, such as ReLU (rectified linear unit), are also part of the architecture. These layers help the network learn more complex patterns. The strengths of CNNs lie in their ability to look at small patches of the input data, thus reducing the number of parameters; and the ability to capture features at different levels of abstraction, through stacked convolutions and pooling layers. The CNN considered in the numerical tasks is detailed in \ref{subsec:cnn}.

\subsection{Interpretability in Machine Learning}
Supervised learning models have demonstrated remarkable capabilities in various domains, including image recognition, natural language processing, and medical diagnosis. However, the inherent complexity of these models often makes it challenging for users to understand how they arrive at their predictions or decisions. This lack of interpretability can be a significant barrier to the deployment of ML systems in real-world applications where transparency and trust are crucial. To aid this point, interpretability refers to the ability to explain and understand the decisions made by ML models in a human-understandable manner. In recent years, there has been growing interest in developing interpretable ML techniques to address this challenge \cite{molnar_interpretable_2022, zhang_survey_2021, ribeiro_why_2016, lundberg_unified_2017, doshivelez_rigorous_2017}. Interpretability can take the form of defaulting to using simpler models which are inherently interpretable. Such models include linear regression, logistic regression, decision trees, naive Bayes classifier and more \cite{molnar_interpretable_2022}. This inherent characteristic simply means that the decision-making process for these methods can be traced and understood. On the other hand, techniques powered by artificial neural networks are notoriously difficult to interpret. This lack of interpretability comes down to their complex nature involving multi-layered matrix calculations which are infeasible to trace. To interpret these techniques, one has to design interpretability (or explainable) methods. Interpretability methods can be an active part of the training process, or take the form of postprocessing once the training has taken place, also known as passive methods. Passive methods are often easier to integrate in many existing ML workflows \cite{molnar_interpretable_2022}. Below, we consider an example of an interpretable method, called LIME, which is the interpreter of choice for our numerics results.

\subsection{Local Interpretable Model-Agnostic Explanations (LIME)}
One of the main interpretable techniques to surge in response to the interpretability challenge is the well-known local interpretable model-agnostic explanations (LIME) \cite{ribeiro_why_2016}. LIME is a model-agnostic method that explains the predictions of any black-box ML model by approximating its behavior with an interpretable local model. It belongs to the class of so-called surrogate models, which utilizes simpler, or inherently interpretable models, to explain more complex ones, such as neural networks. In its anatomy, LIME generates perturbed samples around the instance of interest and observes how the black-box model's predictions change. It then trains an inherently interpretable model, such as a linear regression or decision tree, on these perturbed samples to approximate the local behavior of the black-box model. The LIME framework aims to find an interpretable model $g$ that approximates the behavior of a complex model $f$ locally around a specific instance to be explained $x$, as below:
\begin{equation}
\xi (x) = \argmin_{g \in G} L(f,g,\pi_x) + \Omega(g) \label{eq:lime}
\end{equation}
where weighting function $\pi_x$ is the locality measure which emphasizes the importance of points near $x$. This is achieved by minimizing a loss function $L$ that measures the fidelity of $g$ to $f$ near $x$, while also penalizing the complexity $\Omega$ of $g$ to ensure it remains simple and interpretable. Note that LIME optimizes the loss function $L$, whereas the value of $\Omega$ is to be determined by the user. LIME method is an active subject of study  \cite{alvarezmelis_robustness_2018, slack_fooling_2020, laugel_defining_2018, zhang_why_2019}

\section{Numerical Methods}\label{sec:numerical_methods}
\subsection{Dataset Generation}
A total of $329$ mode multiplexers were designed using SPINS. We used a design region of $3 \, \mu m \times 3 \, \mu m$ and a simulation region of $6 \, \mu m \times 6 \, \mu m \times 2 \, \mu m$. Waveguide widths were set to $0.5 \, \mu m$ for both single mode input waveguides, and $1 \, \mu m$ for the multimode output waveguide. Transmissions were set to $1$ for wg$_{0}$(TE$0$) to wg$_{2}$(TE$0$) (Mode~1) conversion, $1$ for wg$_{1}$(TE$0$) to wg$_{2}$(TE$1$) (Mode~2) conversion, and 0 for other cases. A minimum feature size of $0.08 \, \mu m$ was used. Simulations were performed at a wavelength of $1.55 \, \mu m$. A total of $18$ iterations (fourteen continuous and four discrete) were performed. Eighteen iterations provided a good middle ground between the average device performance and total number of devices.

Following SPINS optimization, broadband spectral response (transmission spectra) of the multiplexers were obtained from $1.45 \mu$m to $1.65 \mu$m using Tidy3D. The devices displayed bandwidths higher than conventional designs, as expected. Therefore, we used a $0.5\,\mathrm{dB}$ bandwidth (i.e.the bandwidth over which the transmission is 0.5 dB less than the peak transmission at 1550 nm) to characterize the devices, rather than the $1\,\mathrm{dB}$ or $3\,\mathrm{dB}$ bandwidths typically employed in conventional designs. A few devices had a large 0.5 dB bandwidth, greater than $200\,\mathrm{nm}$ (which we will take as the high-bandwidth threshold). The mean bandwidths of Mode~1 and Mode~2 were approximately $158\,\mathrm{nm}$ and $141\,\mathrm{nm}$, respectively. For each device, the normalized bandwidth for each mode was calculated by dividing by the mean bandwidth for that mode. Each device was then sorted into two classes based on the sum of its normalized bandwidths: class-$0$ or low-bandwdith class with $110$ devices (normalized sum $\leq1.8$), and class-$1$ or high-bandwidth class with $111$ devices (normalized sum $\geq2.35$). Our final dataset had $221$ mode multiplexers. The remaining devices (with medium bandwidths) were discarded.

\subsection{Convolutional Neural Network (CNN) Architecture}\label{subsec:cnn}
We implemented a CNN model to classify LIME-explained multiplexer images into two categories: high and low bandwidth. The training dataset comprised images from both experimental configurations, with all low-bandwidth images grouped as one class and all high-bandwidth images as the other. This unified dataset enables the CNN to learn bandwidth-related patterns that are consistent across different modes of operation.
\renewcommand{\arraystretch}{1.0}
\begin{table}[h]
\centering
\small
\begin{tabular}{@{} l p{5cm} @{}}
\toprule
\textbf{Hyperparameter} & \textbf{Value} \\
\midrule
Model Type & Sequential CNN \\
Input Shape & $256 \times 256 \times 1$ (Grayscale) \\
Convolutional Layers & 3 layers (32, 64, 128 filters) with L2 regularization \\
Kernel Size & $3 \times 3$ (for all conv layers) \\
Activation Function & ReLU \\
Batch Normalization & After each convolutional layer \\
Pooling Type & Max Pooling ($2 \times 2$ after each conv layer) \\
Dropout & Spatial Dropout (0.3 after first pooling), Dropout (0.4 and 0.3 in FC layers) \\
Global Average Pooling & Yes (replaces flattening) \\
Dense Layers & 128 units and 32 units with ReLU \\
Output Layer & 1 unit with Sigmoid (for binary classification) \\
Loss Function & Binary Crossentropy \\
Optimizer & Adam (default learning rate) \\
Batch Size & 32 \\
Epochs & 20 \\
Train/Test Split & 80/20 \\
\bottomrule
\end{tabular}
\caption{CNN architecture and training hyperparameters.}
\label{tab:cnn-hyperparams}
\end{table}

\begin{figure*}[t]
    \centering
    \includegraphics[width=0.90\linewidth]{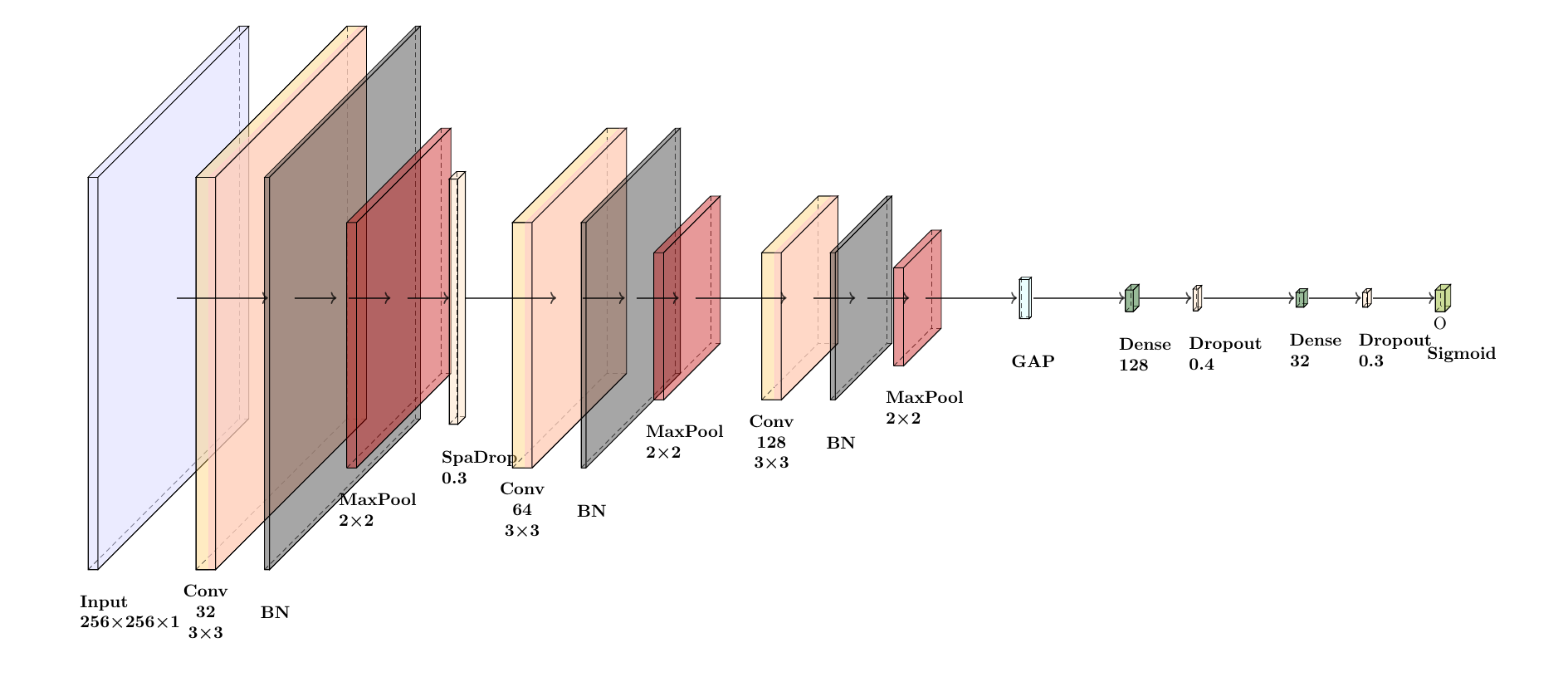}
    \caption{\justifying Architecture of the CNN employed in this study. The network accepts grayscale images of size $256 \times 256$ as input. The model comprises three convolutional layers with filter sizes of $3 \times 3$ and increasing depths of $32$, $64$, and $128$, each followed by batch normalization (BN) and $2 \times 2$ max pooling. A spatial dropout (SpaDrop) layer with a dropout rate of $0.3$ is applied after the first max pooling layer to promote regularization. The convolutional stack is followed by a global average pooling (GAP) layer. The resulting feature vector is passed through two fully connected (dense) layers of $128$ and $32$ units, respectively, with dropout rates of $0.4$ and $0.3$. Finally, a single output node with a sigmoid activation function produces the prediction.}
    \label{fig:cnn}
\end{figure*}

The CNN architecture was intentionally designed to be lightweight, considering the small size of the dataset and the subtle visual distinctions between the two classes (depicted in \Cref{fig:cnn}). This architecture comprises of three convolutional layers with increasing filter depths: $32$, $64$, and $128$, respectively. The network accepts grayscale input images of size $256 \times 256$ and follows a regularized architecture optimized for generalization on limited data. It begins with three convolutional layers using filter depths of $32$, $64$, and $128$, respectively, each with a $3 \times 3$ kernel and ReLU activation. Each convolutional block is followed by batch normalization, $2 \times 2$ max pooling, and spatial dropout to reduce overfitting and encourage feature robustness. Additionally, L2 regularization with a penalty coefficient of $0.01$ is applied to all convolutional layers.

Instead of flattening the output of the final convolutional block, a global average pooling layer is used to minimize overfitting by reducing the number of trainable parameters. The pooled output is passed through two dense layers with $128$ and $32$ neurons, respectively, both using ReLU activation. Dropout layers with rates of $0.4$ and $0.3$ are applied after each dense layer to further mitigate overfitting. The final output layer consists of a single neuron with sigmoid activation, yielding a probability score for binary classification. The model is trained using the Adam optimizer with binary cross-entropy as the loss function \cite{kingma2015adam}. Binary cross-entropy is commonly used for binary classification tasks where the target variable has two classes. It measures the difference between the predicted probability $\hat{y}$ and the actual label $y \in \{0, 1\}$ using the following formulation:
\begin{equation}
\mathcal{L}_{\text{BCE}} = -[y \log(\hat{y}) + (1 - y) \log(1 - \hat{y})]
\end{equation}
This loss function penalizes predictions that deviate from the true label, with greater penalties for confident but incorrect predictions. The function outputs lower values when the predicted probability closely matches the actual class and higher values otherwise. In this work, since the classification task involves distinguishing between two categories --- high and low bandwidth --- binary cross-entropy is an appropriate choice. It allows the network to output a single probability score through a sigmoid activation function in the final layer, reflecting the confidence of the prediction for the positive class.
\subsection{CNN Training Setup}
\begin{figure*}[t]
    \centering
    \subfloat[]{
        \includegraphics[width=0.47\textwidth]{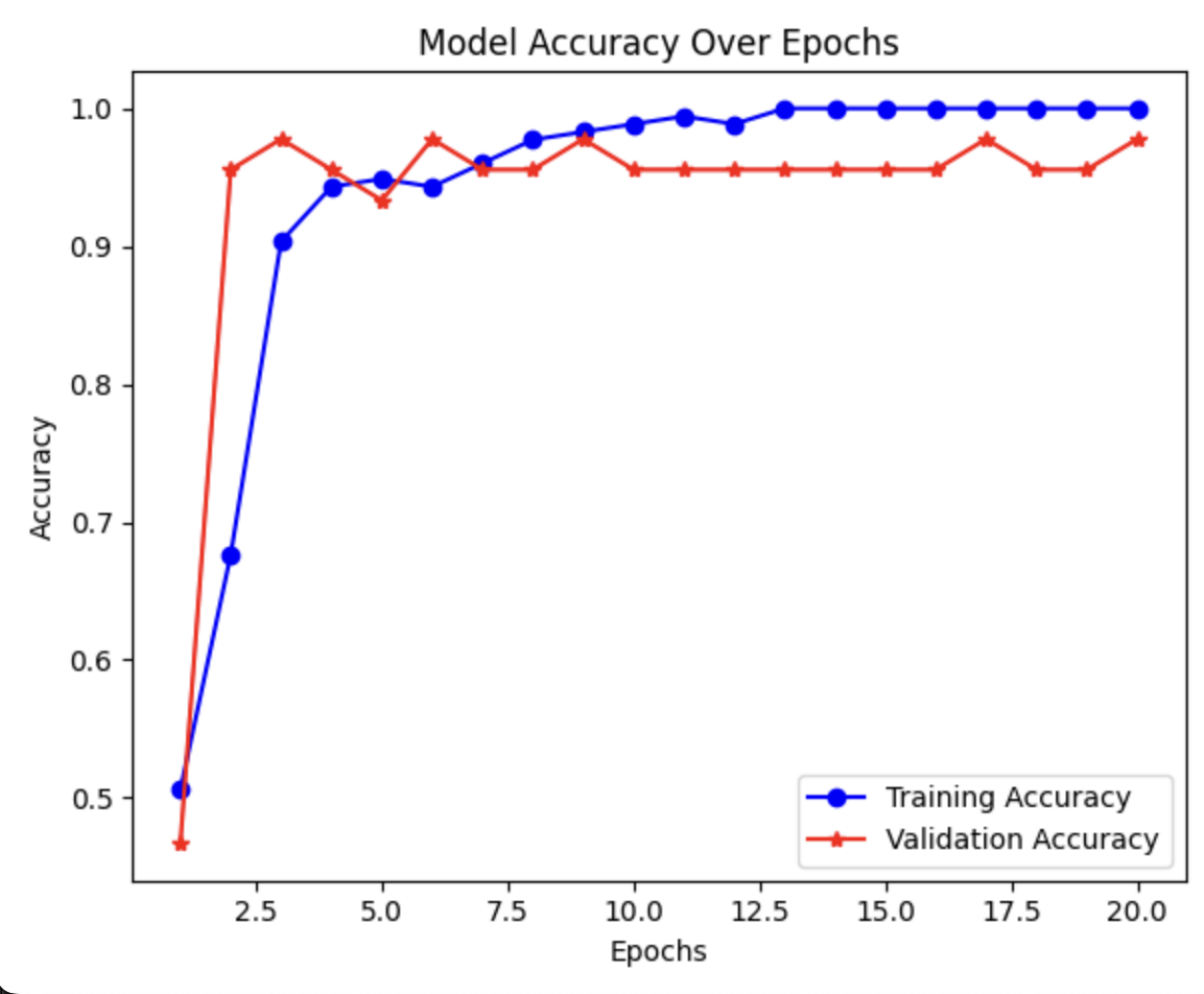}
        \label{fig:training-plot}
    }
    \hfill
    \subfloat[]{
        \includegraphics[width=0.47\textwidth]{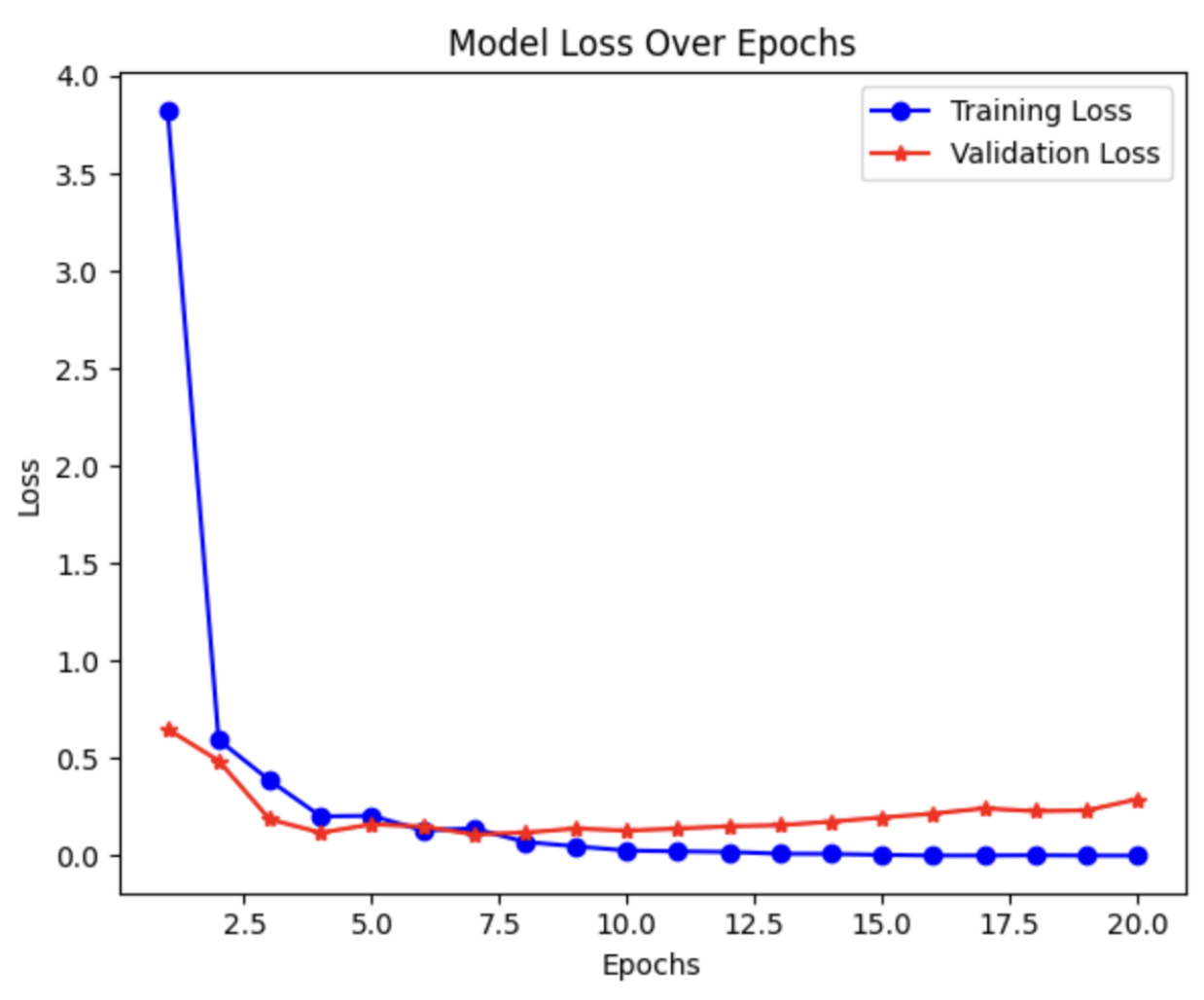}
        \label{fig:loss-plot}
    }
    \caption{Training (a) and Loss (b) curves of the CNN model across the number of epochs.}
    \label{fig:model-plots}
\end{figure*}
To classify LIME-explained multiplexer images into bandwidth categories, we trained a single CNN model using a combined dataset of both experimental configurations: Mode~1 and Mode~2. The input images were grayscale, normalized to the range $[0, 1]$, and resized to $256 \times 256$ pixels. The input images were normalized to the range $[0, 1]$ to ensure numerical stability during training. Normalization helps standardize the input distribution, which accelerates the convergence of the model by preventing large gradient updates. It also makes the optimization process more stable and efficient, particularly when using adaptive optimizers such as Adam. Furthermore, normalizing pixel intensities aligns the scale of input features, enabling better learning dynamics across all layers of the CNN. 

Each image was reshaped into a $(256, 256, 1)$ tensor to explicitly define the presence of a single grayscale channel. Although grayscale images inherently have only one channel, many deep learning frameworks (such as TensorFlow \cite{tensorflow2015} and Keras \cite{chollet2015keras} which are used for our CNN architecture) expect the input to the convolutional neural network to have a 3D tensor shape: \texttt{(height, width, channels)}. Reshaping ensures compatibility with the CNN architecture, which is designed to process inputs with an explicit channel dimension. This also maintains consistency across the data pipeline and allows the model to correctly apply convolutional operations across the spatial and channel dimensions. The performance of this model in terms of the accuracy and loss metric is shown in \Cref{fig:model-plots}.

All low-bandwidth images from both modes were assigned to class $0$, and all high-bandwidth images to class $1$. The dataset was split into training and testing subsets using an $80:20$ ratio, and the labels were encoded as binary integers. The CNN model was trained for $20$ epochs with a batch size of $32$ using the Adam optimizer with default parameters. Given the binary nature of the classification task, the model's final output layer consisted of a single neuron with sigmoid activation, and the loss function used was binary cross-entropy. Accuracy was tracked as the primary evaluation metric. To improve generalization and mitigate overfitting, the model incorporated several regularization techniques, including L2 weight penalties in all convolutional layers, batch normalization after each convolution, spatial dropout following the first pooling layer, and global average pooling instead of flattening. The fully connected layers were followed by dropout layers with rates of $0.4$ and $0.3$, respectively. The final trained model demonstrated steady improvements in training and validation accuracy and was saved in HDF5 format for subsequent evaluation using LIME-based explanations and feature importance analysis.

\subsection{LIME Interpretation Methodology}
\begin{figure}[t]
  \centering
  \includegraphics[width=0.99\linewidth]{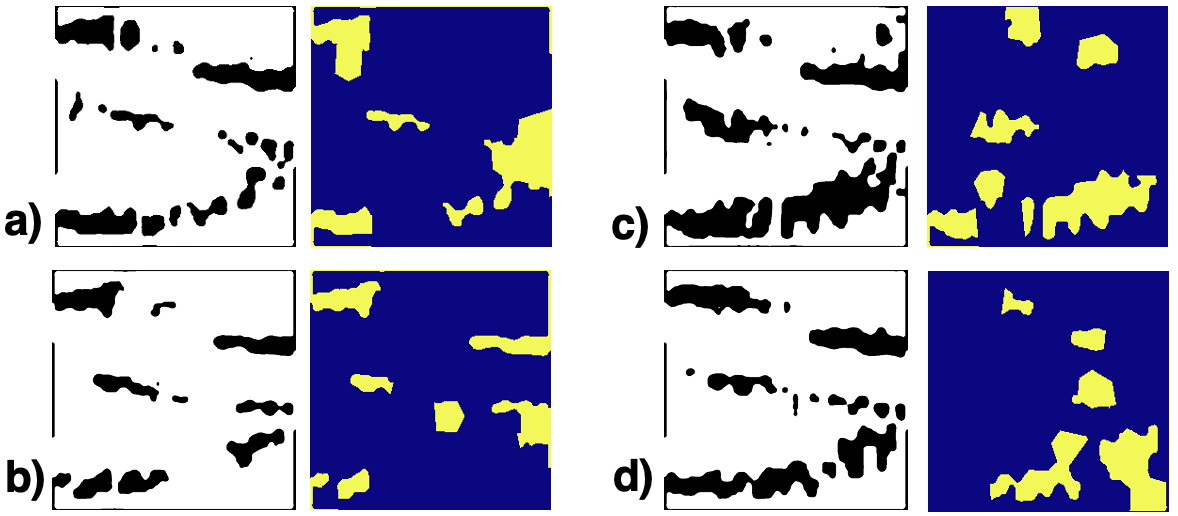}
  \caption{\justifying LIME explanations for Class~0 (low bandwidth) devices (a, b) and Class~1 (high bandwidth) devices (c, d). Each pair displays the original image (grayscale), and the blue-yellow binary mask with important regions in yellow.}
  \label{fig:lime-grid}
\end{figure}

We used LIME to interpret the predictions made by the trained CNN on the multiplexer dataset. Each grayscale image was first converted into a 3-channel RGB format to comply with LIME's input requirements. For each image, LIME generated a set of perturbed samples by selectively masking out image superpixels and observing the model's response to these variations. We used the SLIC algorithm (simple linear iterative clustering) \cite{achanta2010slic} to define superpixels, with $1000$ perturbations per image. The hide-color parameter was set to $255$ to focus on etched regions by controlling the background contrast. The default segmentation commonly used in LIME is quick shift \cite{vedaldi2008quickshift}, which tends to segment images into larger, more regularly shaped regions that may overlook fine details. In comparison, SLIC segments images into smaller chunks, making local density modes more tractable and allowing for more distinct and precise capture of subtle features. The resulting explanations highlight the regions most influential to the model's prediction. For each image, we saved the following: (i) the original (grayscale) image, and (ii) the corresponding binary importance mask from LIME, visualized using a blue-yellow colormap (see \Cref{fig:lime-grid}). The LIME binary mask shows the most important superpixels for a given prediction. The predicted class was used to group and save LIME outputs into class-wise directories.

SLIC segmentation clusters pixels based on both color similarity and spatial proximity in a five-dimensional space, consisting of the CIELAB color space \cite{luo2015cielab}, defined to represent colors by lightness (L), chromaticity axes (a, b), and the spatial coordinates (x,y). To properly combine these distances, SLIC uses a normalized distance measure that balances color difference and spatial distance, with a compactness parameter $m$ controlling the trade-off between adherence to image boundaries and superpixel regularity. Initially, $K$ superpixel centers are sampled at regular grid intervals and moved to nearby positions of low image gradient to avoid unstable regions. Each pixel is then associated with the nearest center based on the defined distance measure, and the cluster centers are iteratively updated until convergence. Connectivity enforcement is performed as a post-processing step to ensure spatial coherence.

As the dataset consisted of grayscale images, we first replicated the single-channel images across three channels to form pseudo-RGB inputs. When applying SLIC, the RGB images were internally converted into the CIELAB color space. Since the R, G, and B values were identical across channels, the resulting LAB representation exhibited near-zero values for the chromaticity components (a and b), meaning that pixel variation was dominated by differences in lightness (L) rather than hue or saturation. As a result, SLIC clustering in our case was  guided by brightness and spatial proximity alone. This behavior was particularly well-suited to our dataset, where fine, localized etched patterns --- rather than color variations --- are distinguished between classes.

We set the number of superpixels to $100$, with a compactness value of $10$, and enforced spatial connectivity to improve the consistency of the segmented regions. When applying LIME, we focused exclusively on highlighting positive features that contributed towards the model's predicted class and used a hide color value of $255$ to mask out non-informative areas, thereby maintaining the visual context of the etched regions. By tuning the segmentation parameters appropriately, we ensured that small but meaningful class-specific features were preserved without overwhelming the model with unnecessary noise. Moreover, using superpixel-level perturbations, set to $1000$ in our case, we significantly reduced the computational burden in LIME while still allowing precise and interpretable explanations.

\section{Numerical Results}\label{sec:numerical-results}
\subsection{Analysis of LIME Interpretations}

\begin{figure*}[ht]
  \centering
  \includegraphics[width=0.8\textwidth]{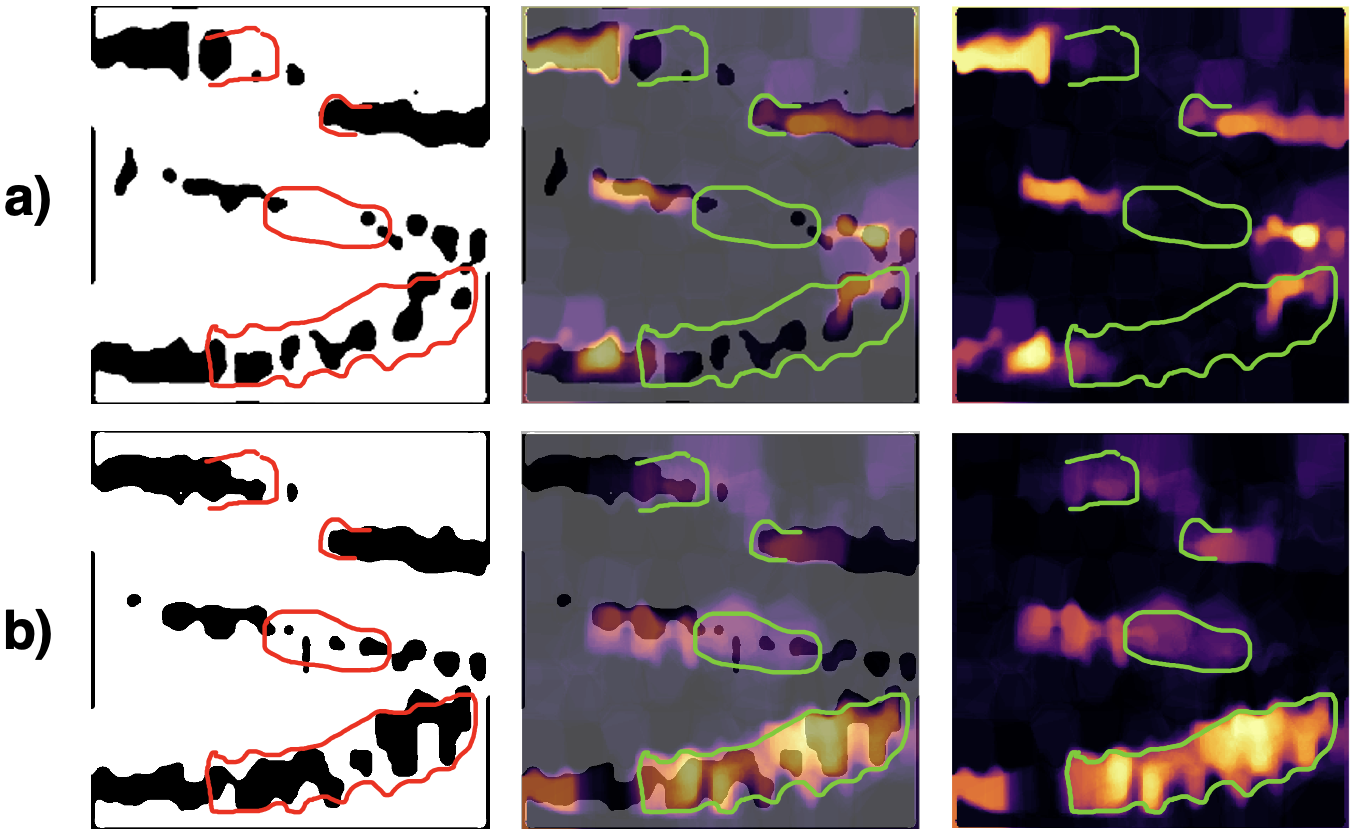}
  \caption{Multiplexer M1 from low-bandwidth class (a), and M4 from high-bandwidth class (b). From left to right: the device, LIME heatmap for the corresponding class superimposed on the device, LIME heatmap. Highlighted (red/green) regions point to differences between both heatmaps and the associated features on device.}
  \label{fig: lime_heatmaps}
\end{figure*}

The LIME masks were averaged to find heatmaps for each class. Each heatmap was superimposed on top of a multiplexer (M1 and M4) from the corresponding class, as shown in \Cref{fig: lime_heatmaps}. These heatmaps highlight features that LIME suggests are important to classify into high-bandwidth or low-bandwidth classes. The circled regions indicate the major differences between both heatmaps. For high-bandwidth devices like M4, LIME seems to be looking for continuous etched regions; while M1 and other low-bandwidth devices seem to favor having more gaps between structures. Note that the highlighted features are at the boundaries of mode propagation. Lower connectivity for low-bandwidth devices means that there are more leaks and small features, both of which could contribute to wavelength-dependent insertion loss. Small features can cause scattering that is wavelength-sensitive and the gaps could cause some wavelengths to leak during transmission. Our LIME heatmaps seem to suggest that these are the main predictors of low bandwidth. 

\subsection{FDTD Simulations in Support of LIME Results}
\begin{figure*}[ht]
  \centering
  \includegraphics[width=\textwidth]{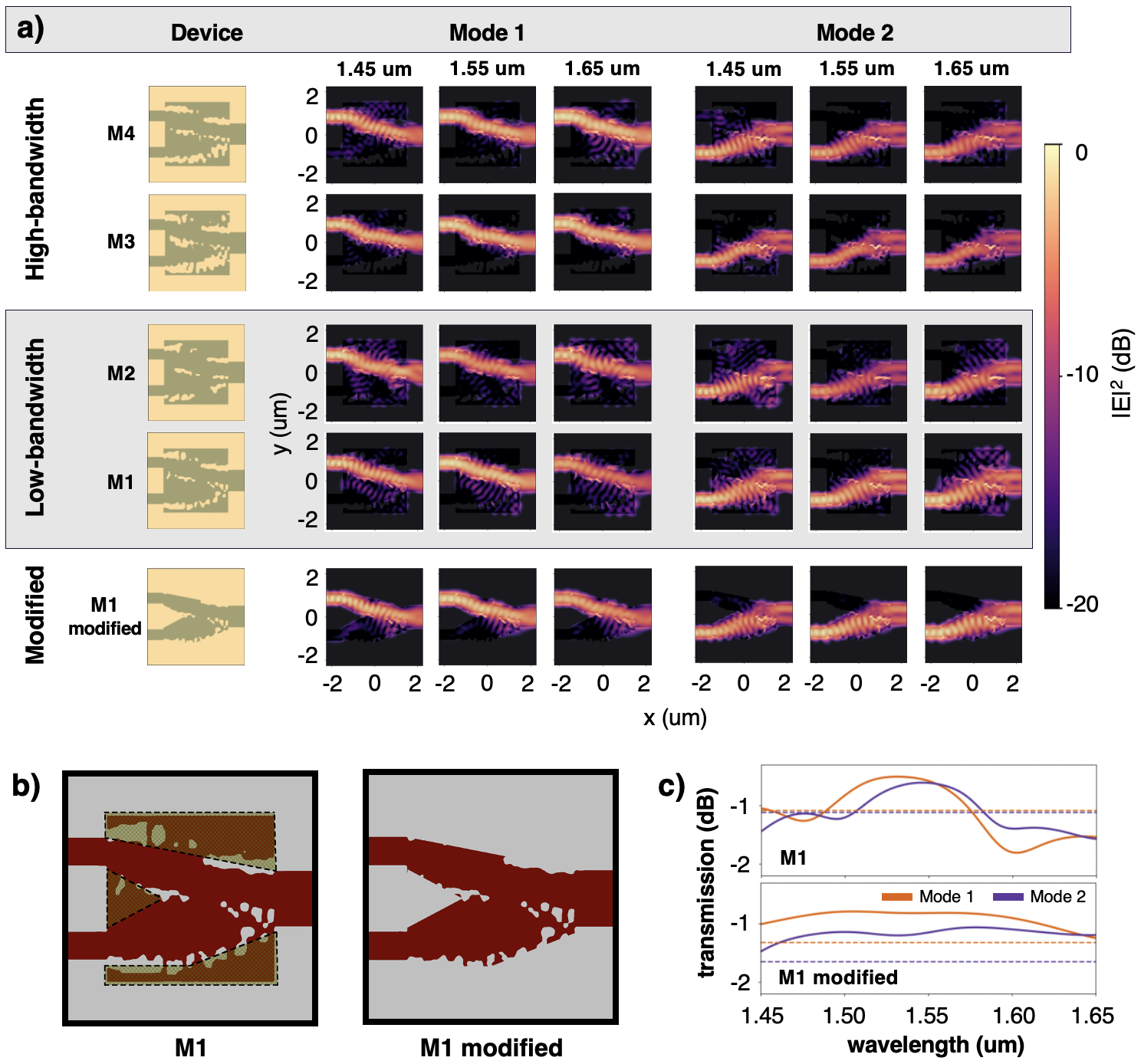}
  \caption{a) The normalized electric intensity fields for devices M4 to M1 (high to low bandwidth) and for the modified M1 (for Modes~1 and 2, and wavelengths $1.45$-$1.65$ $\mu$m). Logarithmic scale is used to highlight weak fields scattering/leaking through the small features in low-bandwidth devices. b) The coarse modifications made to device M1 (shaded regions were cropped), and the transmission spectra of both devices (c). Bandwidths improved despite the rough adjustments and additional loss.}
  \label{fig:fdtd_1.45}
\end{figure*}

We used FDTD simulations to further test our LIME interpretation. \Cref{fig:fdtd_1.45}(a) shows the normalized electric intensity fields for devices M4 to M1 (high to low bandwidth). The fields are plotted on logarithmic scale to highlight weak fields. Field propagation at wavelengths $1.45$ $\mu$m, $1.55$ $\mu$m, and $1.65$ $\mu$m is included for both modes. Going away from the central wavelength, the low-bandwidth devices show more stray fields leaking through the gaps. This observation is in agreement with the LIME heatmap analysis. 

Furthermore, we made some coarse adjustments to the low-bandwidth device by removing some regions as shown in \Cref{fig:fdtd_1.45}(b). These adjustments were intended to improve connectivity between structures by reducing small features and gaps, essentially giving it more high-bandwidth characteristics based on our LIME analysis. Note that such coarse modification of an inverse designed component is sure to affect performance --- our aim was to check if the bandwidth would manage to increase despite this loss. At $1.55$ $\mu$m, the transmission decreased from $-0.58$ dB to $-0.82$ dB for Mode~1, and from $-0.61$ dB to $-1.15$ dB for Mode~2, as expected. However, other wavelengths were not similarly affected; the bandwidth of M1 was $84$ nm for Mode~1, and $76$ nm for Mode~2. After modification, both bandwidths crossed the $200$ nm threshold. The electric intensity fields of the modified device also show reduced scattering at $1.45$ and $1.65$ $\mu$m. Thus, the LIME-informed coarse adjustments enhanced the bandwidth despite an increase in insertion loss. The next step was to use these insights to improve bandwidth without loss in transmission.

\section{Application: LIME-informed Initial Condition for Inverse Design}\label{sec:initial-condition}
\begin{figure*}[ht]
  \centering \includegraphics[width=\textwidth]{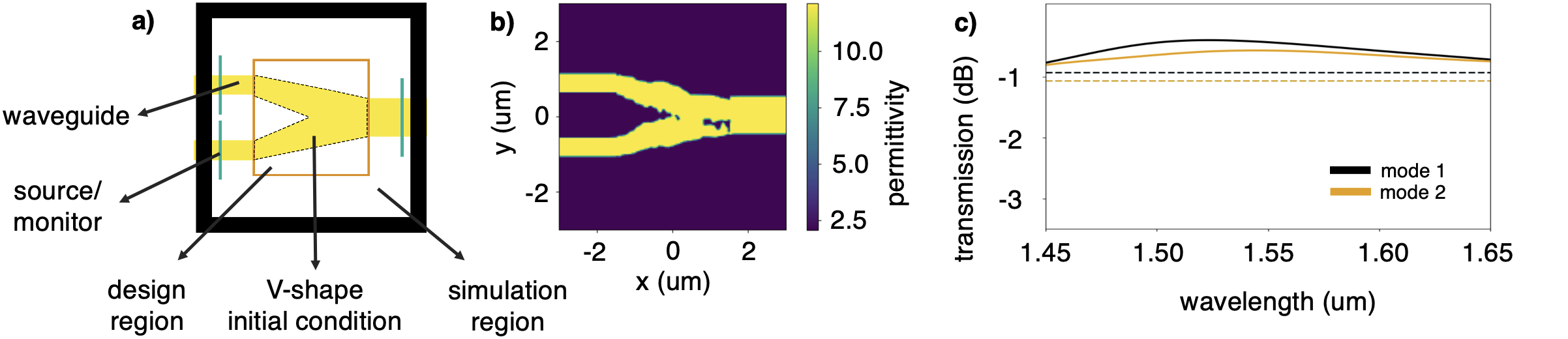}
  \caption{LIME-informed initial condition for inverse design (a), and the final device (b). The transmission spectra (c) shows the device has a $0.5$ dB bandwidth (dotted lines) above $200$ nm for both modes.
  }
  \label{fig:initial-condition}
\end{figure*}
Similarly to most machine learning algorithms, the choice of the initial condition plays an important role in inverse design optimization in terms of final device performance. SPINS and other packages have generic initial designs to choose from; however, those need not be the best suited for improving a desired characteristic of a particular device.
Equipped with the insights gained from our LIME interpretations, we chose a simple V-shaped structure as shown in \Cref{fig:initial-condition}(a) as our initial design in hopes of improving the bandwidths of both modes. Since there were no gaps along propagation, such a basic structure was hypothesized to reduce the leaky nature and scattering effects across different wavelengths while still giving ample space for SPINS to convert it into a functional mode multiplexer. 

All design parameters were kept similar to our generated dataset, except for the initial condition. A total of $20$ devices were designed. One of the resulting devices is given in \Cref{fig:initial-condition}(b). The structure is noticeably different from those in the previous dataset; it is quite similar to the V-shape that was used as our initial condition. This indicates that the chosen initial condition has likely altered the optimization route and is providing a different family of solutions. The spectral response of the devices were analyzed using the FDTD solver. The LIME-informed devices showed higher bandwidths, with all twenty of them crossing the $200\,\text{nm}$ threshold for $0.5$ dB bandwidth across both modes. In contrast, only seven of the $329$ multiplexers in our original dataset crossed the $200$ nm bandwidth threshold for both modes. The initial condition also improved insertion losses at $1.55 \mu$m, with the lowest transmission being $-0.43$ dB for Mode~1 and $-0.81$ dB for Mode~2, whereas those for the original dataset were $-1.16$ dB and $-1.53$ dB, respectively (two outliers in the original dataset that showed extremely poor transmission were not considered).

\section{Application: Design and Classification Considerations for Multiplexers}\label{sec:mux-generation}
\begin{figure}[ht]
    \centering
    \includegraphics[width=0.99\linewidth]{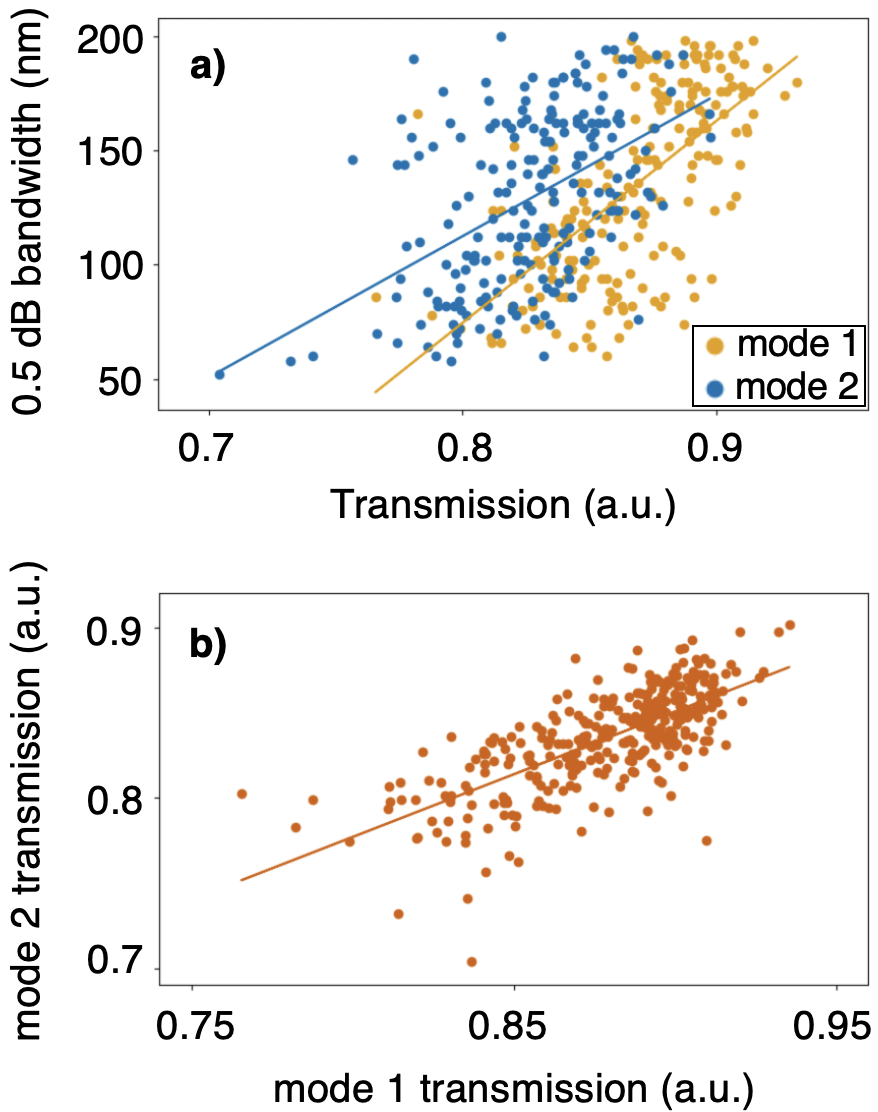}
    \caption{Scatter plots for a) 0.5 dB bandwidth vs transmission at $1.55~\mu$m (R is $0.65$ for Mode~1, and $0.49$ for Mode~2), and b) transmission of Mode~1 vs Mode~2 at $1.55~\mu$m (R $= 0.72$)}
    \label{fig:corr-plots}
\end{figure}
The decision to classify the devices into only two categories was guided by both practical considerations and the structure of the dataset. We deemed unnecessary to introduce additional classes --- such as one for devices with high bandwidth in Mode~1 but low in Mode~2, and another for the reverse. Such configurations lack practical relevance, as real-world applications rarely require high bandwidth in only one operational mode while tolerating poor performance in the other. Typically, both modes are expected to maintain high bandwidth performance.

In addition, the limited size of the dataset constrained the viability of further subclassification. Adding more categories would reduce the number of data points per class, thereby undermining the reliability of both classification and subsequent interpretation. There were also different correlations among device characteristics like bandwidth and transmission, which could have made investigating additional classes more difficult. For instance, \Cref{fig:corr-plots} (a) shows a moderate to weak correlation between bandwidth and transmission efficiency at $1.55 \mu$m (Pearson's correlation coefficient (R) is $0.65$ for Mode~1 and $0.49$ for Mode~2). A potential reason could be that the features suggested by LIME to favor higher bandwidth --- such as fewer gaps --- might also improve transmission at the central wavelength. For this plot, we removed devices that crossed the $200\,\text{nm}$ bandwidth threshold since their exact bandwidths are unknown. Two outlier devices with extremely poor transmissions ($< -8$ dB) were also removed from the plot. There is also some positive correlation between the transmissions for both modes (R $= 0.72$, \Cref{fig:corr-plots} (b)). We believe this to be a consequence of the optimization process --- devices that meet higher transmission in one mode at the start of the optimization might place a stronger demand on improving the transmission of the other mode to satisfy the objective function. These insights are interesting and can be part of further exploration. For this work, these observations support the use of a single, combined bandwidth metric for classification.

\section{Conclusion \& Future work}\label{sec:conclusion}
In this study, we have explored the integration of interpretable machine learning techniques into the inverse design process for photonic chip design. More specifically, we have considered the LIME interpreter applied to mode multiplexers, an important component in photonic circuits. The results presented here are relevant beyond integrated photonics and into broader topological optimization strategies, materials design, and other fields where complex systems are optimized. The integration of interpretable ML methods into photonic design workflows is important for both fields. From the photonics points of view, the results help engineers better understand the relationship between design parameters and device performance, thus leading to more precise and efficient photonic circuit design. From an ML perspective, this work contributes to ongoing initiatives and development of interpretable ML in real-life scenarios in engineering.

We recognize the small-scale sample size of our study as a limitation. The results presented demonstrate the potential of our approach as a proof-of-concept idea, demonstrating the feasibility of applying our approach to the specific problem at hand. However, the insights gained from this study already provide a starting point for advancing the use of interpretable machine learning in photonic chip design and other engineering disciplines such as aerospace engineering, material science, or electrical engineering.

Moving forward, we suggest several avenues for further research. First, more complex photonic components such as multimode beam splitters, couplers, filters, interferometers, and modulators can be studied, and different performance metrics such as insertion loss and fabrication sensitivity may be considered using interpretability techniques. Secondly, a wider range of interpretability techniques can be considered, including saliency maps, other feature attribution methods such as SHAP, and surrogate methods such as decision trees. In addition, there is a need to develop more robust and scalable interpretable machine learning methods tailored specifically to the challenges of inverse design, which are then applied in the design of photonic components. In addition to the work we have presented here, a crucial next step is to actually fabricate the designs informed by LIME and check that they indeed lead to large bandwidths. It is also an opportunity to assess the scalability of LIME-informed designs, ensuring that they can be fabricated efficiently on scale without losing performance due to manufacturing imperfections or material limitations.

\vspace{0.4cm}
\emph{Acknowledgments:} LP thanks Serge Massar for discussions on photonic machine learning and its practical feasibility. Authors thank Jamika Ann Roque, Chris Ferrie, Dening Luan and Patrick Rebentrost for discussions. This work is partially supported by the National Research Foundation, Singapore through the National Quantum Office, hosted in A*STAR, under its Centre for Quantum Technologies Funding Initiative (S24Q2d0009), and by the National Research Foundation, Singapore and A*STAR under its Quantum Engineering Programme (NRF2021-QEP2-02-P05), and by EQUS (CE170100009). LP acknowledges support from an Australian Research Council Centre of Excellence for Engineered Quantum Systems (EQUS) Collaboration Grant awarded for a visit to the Romero group at the University of Queensland, where the idea for this project originated.

\bibliographystyle{apsrev4-2}
\bibliography{bibliography}

\end{document}